# Direct Imaging Revealing Halved Ferromagnetism in Tensile-Strained LaCoO$_3$ Thin Films

Qiyuan Feng[1,2*], Dechao Meng[1,2,3*], Haibiao Zhou[1,4], Genhao Liang[1,2], Zhangzhang Cui[1,2], Haoliang Huang[2,5], Jianlin Wang[2,5], Jinghua Guo[6], Chao Ma[7], Xiaofang Zhai[1,2†], Qingyou Lu[1,4†] and Yalin Lu[1,2,5†]

[1]*Hefei National Laboratory for Physical Sciences at Microscale, University of Science and Technology of China, Hefei, 230026, China.*

[2]*Synergy Innovation Center of Quantum Information and Quantum Physics, University of Science and Technology of China, Hefei, Anhui 230026, China.*

[3]*Microsystem and Terahertz Research Center & Institute of Electronic Engineering, China Academy of Engineering Physics, Chengdu, 610200, China.*

[4]*Anhui Key Laboratory of Condensed Matter Physics at Extreme Conditions, High Magnetic Field Laboratory and Hefei Science Center, Chinese Academy of Sciences, Hefei, 230031, China.*

[5]*National Synchrotron Radiation Laboratory, University of Science and Technology of China, Hefei, Anhui 230026, China.*

[6]*Advanced Light Source, Lawrence Berkeley National Laboratory, Berkeley, California 94720, USA.*

[7]*College of Materials Science and Engineering, Hunan University, Changsha 410082, China*

*These authors contributed equally to this work.

†Corresponding authors: xfzhai@ustc.edu.cn (X.Z.); qxl@ustc.edu.cn (Q.L.); yllu@ustc.edu.cn (Y.L.)

## Abstract

The enigma of the emergent ferromagnetic state in tensile-strained $LaCoO_3$ thin films remains to be explored because of the lack of a well agreed explanation. The direct magnetic imaging technique using a low-temperature magnetic force microscope (MFM) is critical to reveal new aspects of the ferromagnetism by investigating the lateral magnetic phase distribution. Here we show the experimental demonstration of the rare halved occupation of the ferromagnetic state in tensile-strained $LaCoO_3$ thin films on $SrTiO_3$ substrates using the MFM. The films have uniformly strained lattice structure and minimal oxygen vacancies (less than 2%) beyond the measurement limit. It is found that percolated ferromagnetic regions with typical sizes between 100 nm and 200 nm occupy about 50% of the entire film, even down to the lowest achievable temperature of 4.5 K and up to the largest magnetic field of 13.4 T. Preformed ferromagnetic droplets were still observed when the temperature is 20 K above the Curie temperature indicating the existence of possible Griffiths phase. Our study demonstrated a sub-micron level phase separation in high quality $LaCoO_3$ thin films, which has substantial implications in revealing the intrinsic nature of the emergent ferromagnetism.

# I. INTRODUCTION

The correlated oxide thin films and their interfaces represent a rich platform in exploring novel magnetic phenomena, which are absent in their bulks [1-7]. The direct imaging of magnetism in the lateral direction has played a key role in resolving the underlying physics of some extraordinary magnetic phenomena, such as the colossal magnetoresistance effect [8-12]. Very recently, the scanning superconducting quantum interference device (SQUID) microscopy has revealed lateral magnetic distributions in the $LaAlO_3/SrTiO_3$ two-dimensional electron gas [13] and ultrathin $LaMnO_3$ films [14, 15]. The magnetic force microscopy (MFM) has an advantage over the scanning SQUID by being able to operate under much higher fields. Recent improvements of the force measurement technique have warranted the high accuracy in charactering small signals of ultrathin films [16, 17]. Thus it is particularly privileged in exploring magnetic phase distributions of ultrathin films under high magnetic fields.

The tensile-strained $LaCoO_3$ ultrathin film is a puzzling example of emergent magnetism in correlated oxide heterostructures. The $LaCoO_3$ single crystal exhibits a diamagnetic low-spin state (S=0) below 90 K, above which a small amount of Co atoms are thermally activated to the intermediate-spin state (S=1) or high-spin state (S=2) and the system transits to a paramagnet [18-20]. When $LaCoO_3$ thin films are tensile strained, it surprisingly turns into a strong ferromagnet with a Curie temperature (Tc) as high as 85 K [21, 22]. It is generally accepted that the tensile strain is necessary to induce the ferromagnetism (FM) while the compressive strain is not much effective [23, 24]. Alternatively, some scanning transmission electron microscopy (STEM) studies have found densely ordered stripes of oxygen vacancies and suggested the correlation between the FM and the oxygen vacancy ordering [25-27]. However, both soft and hard X-ray absorption measurements prescribed an upper limit of oxygen vacancies that is significantly lower than those suggested by STEMs [28-32]. More recent STEM studies reported that stripes can be induced by elongated exposures to electron beams [33, 34]. There are also suggestions that the FM is correlated to the ferroelasticity [30, 32] or the octahedral rotation pattern change [35]. To date, the origin of the FM in LaCoO$_3$ films is still being highly debated. Nevertheless, almost all experiments in high quality LaCoO$_3$ films on SrTiO$_3$ substrates demonstrate agreeable magnetic properties: a saturation magnetization of about 1 $\mu_B$/Co and a Curie temperature of about 85 K. The existing theories primarily assume simple magnetic distributions, such as the uniform Co$^{3+}$ intermediate spin state [36] or the Co$^{3+}$ high spin - Co$^{3+}$ low spin checkerboard state [37, 38]. The former leads to a conducting state which is contrary to the experiments, while the latter leads to a saturation magnetization that is twice of the experimental value. It is noted that nanoscopic phase separations have been observed in hole-doped La$_{1-x}$Sr$_x$CoO$_3$ films [39, 40]. Although an MFM study has been conducted in LaCoO$_3$ films with substantial amount of structural defects [41], it has not been done in high quality films. Therefore, a magnetic phase distribution study of the high quality

LaCoO$_3$ films is necessary, which may shine a new light toward resolving the complicated nature of the emergent FM.

Here we directly imaged the FM state of a high quality LaCoO$_3$ film on a (0 0 1) oriented SrTiO$_3$ substrate with an ultrathin thickness of 11.3 nm (30 unit cell) using a low temperature high magnetic field MFM. Film fabrication and MFM measurement details are available in the Appendix. We performed MFM experiments in a large range of temperature and under high magnetic fields. We found the FM phase coexisting with the non-magnetic phase down to the lowest measured temperature of 4.5 K, even under a magnetic field as high as 13.4 T. The observed FM regions maximally occupy 50% of the film. And the FM regions have the granular shape with typical sizes between 100 nm and 200 nm. We observed preformed ferromagnetic droplets when the temperature is 20 K above T$_C$ which indicates the existence of a Griffiths phase. Our study demonstrated a strong sub-micron level magnetic phase separation in high quality LaCoO$_3$ films with uniform lattice structure and good stoichiometry, which has not been reported before.

## II. RESULTS AND DISCUSSION

The STEM high-angle annular dark-field (HAADF) image of the 30 unit cell LaCoO$_3$ film on the SrTiO$_3$ substrate shown in Fig. 1(a) demonstrates the abrupt interface between the film and the SrTiO$_3$ substrate. Clear Laue fringes around the X-ray diffraction (XRD) (0 0 2) peak in Fig. 1(b) indicate sharp interfaces and smooth surfaces of the high quality film and the out-of-plane lattice constant $c$ is determined to be ~ 3.77 Å. The XRD reciprocal space maps (RSM) around the (0 -2 4), (2 0 4), (0 2 4) and (-2 0 4) peaks are shown in Fig. 1(c), demonstrating that the 30 unit cell film is fully strained to the substrate. The four peaks all exhibit the same reciprocal lattice constant along the reciprocal $L$ direction.

The Co valence of LaCoO$_3$ films is measured by X-ray absorption spectroscopy (XAS) and X-ray absorption fine structure (XAFS). The XAS was performed using the total electron yield (TEY) method, and XAFS was performed using the total fluorescence yield (TFY) method. The TEY is surface sensitive while the TFY detects the entire film thickness [42]. The measured Co valences with experimental uncertainties are shown in Fig. 1(d). Specifically, Co valence of the 30 unit cell film is determined to be (2.98±0.02) and (2.98±0.04) by the XAS TEY and XAFS TFY respectively. The in-plane and out-of-plane magnetic hysteresis loops (M-H) measured at 5 K are shown in Fig. 2(a), and the temperature-dependent magnetization (M-T) curves are shown in Fig. 2(b). The saturation magnetization (~1 $\mu_B$/Co) and the Curie temperature (85 K) agree to previous experimental results [21-27]. The easy axis lies within the in-plane. The coercive fields are 0.21 T and 0.56 T, and the saturation fields are 1.2 T and 2 T along the in-plane and out-of-plane directions, respectively. The exchange bias measurements were performed to investigate possibilities of antiferromagnetic inclusion in the ferromagnetic matrix. The ±3 T field cool MH measurements along both the in-plane and out-of-plane directions are shown in Fig. 2 (c) and (d). The 3 T field is used since it is above the saturation fields in both the in-plane and out-of-plane directions. None observable exchange biases are found in both directions, which do not support the possibility of coexisting antiferromagnetism.

STEM electron energy loss spectra (EELS) measurements were carried out to study possible valence variations at different depth of the film. The results are shown in Fig. 3. In Fig. 3c and 3d, the surface, middle and interfacial blocks of the 30 unit cell film, each of about 10 unit cells, exhibit similar O $K$-edge and Co $L$-edge spectra. The size of the oxygen pre-peaks, the energy of Co $L_{3,2}$ edges and the Co $L_3/L_2$ peak ratio are all similar in the three blocks at different depth of the film. As a comparison, the EELS results of a 7 unit cell film with large concentrations of oxygen vacancies are also added in Fig. 3. Thus the STEM-EELS results preclude any significant vertical valence variation in the 30 unit cell film and the local valence of Co is close to +3, agreeing to the macroscopic XAS and XAFS results. Furthermore, it is known that the lattice constant could also serve as a fingerprint feature of local oxygen vacancy concentration and Co valence [43]. The lattice constants are calibrated by STEM-HAADF images measured along both the in-plane and out-of-plane directions, as shown in Fig. 4. The in-plane lattice constant $a$ is always around 3.9 Å. The out-of-plane lattice constant $c$ in the first 2 to 3 unit cells is close to that of $SrTiO_3$, probably related to the interface octahedral rotation pattern change. Above the 2-3 interfacial layers, the rest of the film does not show an obvious trend of lattice constant change. The vertical lattice constant is calibrated to be ~3.8 Å, which agrees to the overall $c$ of 3.77 Å measured by XRD. The HAADF images demonstrate a fairly constant lattice at different depth within experimental uncertainty. Therefore, the XAS, XAFS, microscopic STEM-EELS and HAADF results all demonstrate the high-quality of the 30 unit cell LaCoO$_3$ film both structurally and stoichiometrically.

We performed MFM imaging measurements to study the magnetization saturation process of the film at 4.5 K with applied perpendicular magnetic fields as large as 13.4 T, and we found striking evidences of FM regions coexisting with non-FM regions. Atomically-flat surfaces with clear 1 unit cell high terraces are observed in Fig. 5(a). Figs. 5(b)-(h) show the MFM images recorded during the field increasing after zero field cooling. The different colors in the MFM image indicate different interactions between the tip and the local area. Since the MFM tip is only sensitive to the out-of-plane magnetization, the yellowish color represents negligible interactions from weak/non-FM or in-plane magnetized FM regions [44], and bluish color represents attractive interactions from FM regions. As shown in Fig. 5(b), the MFM image measured at 0 T shows well separated FM droplets (blue color) immersed either in weak/non-FM regions or in-plane magnetized FM regions. With the field increasing, the FM droplets gradually expand [Fig. 5(c)-(e)]. For field above 3 T, the size of the detected FM regions and the contrast between the FM and the other regions remain almost the same [Fig. 5(e)-(h)]. The saturation field in the MFM measurement is consistent with that of the bulk SQUID result measured along the out-of-plane direction. It thus excludes the possibility of the other regions (yellowish color) being FM along the in-plane direction. Another important observation is that, even under the maximal field of 13.4 T, the FM phase only occupies about 50% of the entire area [Fig. 5(h)]. The non-FM phase could be paramagnetic or diamagnetic (spin singlet) since $LaCoO_3$ has a very rich magnetic phase diagram in the bulk [19].

The quantitative analyses of the above magnetization saturation process are shown in Figs. 5(i)-(k), in which the area percentage and the size distribution of the FM regions are given. Right after cooling to 4.5 K, the FM area percentage is only ~ 17% [Fig. 5(b)], and it increases to ~ 50% when the field is increased to 3.0 T. Since the detected MFM signal is the shift of the resonance frequency, the root mean square (RMS) of the signal very accurately demonstrate the FM phase development tuned by the applied field. As shown in Fig. 5(j), the RMS reaches to a saturation level when the field is increased to 3 T, above which the RMS slightly decreases. The subsequent slow and nearly linear decrease of the RMS with field increasing from 3.0 to 13.4 T indicates that magnetic moments in the non-FM regions are slowly aligned by the large external field. Fig. 5(k) shows the statistic histogram of the size distribution of FM regions. And the most abundant FM regions have sizes of about 160 nm, which is more than one order of magnitude larger than the thickness.

Next, we performed MFM imaging during the magnetization flipping process at a higher temperature of 45 K, in order to avoid the low-temperature paramagnetic background of the substrate. The results demonstrate that the reversal process most likely occurs between 0.1 T and 1 T. The saturation field is about 2.3 T and the statistics of the FM phase distribution are both similar to those found at 4.5 K. The film was cooled under an applied field of 2.3 T to saturate the FM phase. Right after the field cooling [Fig. 6(a)], non-FM regions (yellow color) are observed coexisting with the FM regions (blue color). As the field reducing from 2.3 T to 0 T, the MFM image contrast becomes weaker and weaker, but the image pattern is barely changing. Then a field of opposite direction was applied to initiate the magnetization reversal process. The MFM image taken at -0.1 T [Fig. 6(e)] is almost the same as that taken at 0 T [Fig. 6(d)], indicating that such a small field is not enough to reverse the magnetization. Dramatic changes are observed when the field is decreased to -0.5 T: the contrast becomes larger and the pattern exhibits a large change. Further changes of the image pattern and enhancements of the contrast can be observed when the field is decreased to -1 T. When the field is decreased to -5 T, both the pattern and the contrast of the image are almost identical to those taken at 2.3 T. The statistical analyses of the magnetization flipping process are shown in Fig. 6(i)-(k), which demonstrates that the sizes of FM regions after saturation are predominately between 100 nm and 200 nm with a peak near 175 nm. It means that the observed sub-micrometer scaled phase separation is different from the nanoscopic phase separation found in the previous STEM-EELS work in the Sr doped lanthanum cobaltate oxides.

Temperature dependences of the MFM images are measured from 45 K to 105 K, across the Curie temperature of 85 K. The results further confirm the static FM and non-FM phase separation, and also surprisingly demonstrate preformed FM domains well above the Curie temperature. Figs. 7(a)-(d) and Figs. 7(e)-(h) show two series of MFM images measured at fields of 0.3 T and 2.3 T, respectively. As the temperature is increasing, the image pattern is unperturbed but the contrast is gradually decreased. For the 0.3 T field and at 85 K, the contrast became very weak but isolated FM (blue color) regions can still be observed as shown in Fig. 7(c). For 0.3 T and at 105 K, there are almost negligible hint of FM droplets [Fig. 7(d)], which agrees to the zero magnetization above 85 K from the bulk SQUID measurement [Fig. 2(b)]. However, for the 2.3 T applied field, distinct FM and non-FM phase contrast can be observed at 85 K and even 105 K. The measurements demonstrate that the spontaneous transition to long range FM order occurs at 85 K, agreeing to the bulk SQUID measurement. But the MFM measurements show that the short range FM order occurs at a much higher temperature than the Curie temperature. Previously local FM ordered regions were rarely directly measured above the Curie temperature. Theoretically, such occurrence of short range FM orders in the temperature range of $T_G>T>T_C$ was ascribed to the Griffiths singularity due to the random disorder [45, 46]. $T_G$ is the intrinsic ordering temperature in the system free of disorders. Random A or B site disorders which are the typical causes of Griffiths phases in doped manganite oxides [46-48], which is not quite the case in $LaCoO_3$ films. Other possible sources of disorders may include few available amount of oxygen vacancies, although too less to be identified by the XAS, XAFS and EELS.

Typically, when multi domain structures appear in strained thin films, there are usually slight symmetry reductions and the symmetry equivalent XRD peaks would appear at different positions in the reciprocal space. The observation of identical peak positions indicates that the 30 unit cell film exhibits a single domain structure. As a comparison, Fig. 8 shows that a 40 unit cell film exhibits peak position variations along the $L$ direction. With the thickness further increased to 90 and 180 unit cells, cross-hatch-line grain boundaries that are absent in 30 unit cell films are found in the atomic force microscopy (AFM) and scanning electron microscopy (SEM) results shown in Fig. 9. As shown in Fig. 9 (d) and (e), the similarity between the MFM and surface morphology of the 180 unit cell film confirm that the grain boundaries cause the magnetic phase separation in thick films. In contrast, the AFM and SEM images of the 30 unit cell film show flat surfaces with single-unit-cell height terraces. Thus the observed phase separations in single-domain thin films and multi-domain thick films are of different origins, with the former being an electronic effect and the latter being a structural effect.

## III CONCLUSION

To summarize, we successfully grew tensile-strained $LaCoO_3$ ultrathin films with high structural and stoichiometric qualities by pulsed laser deposition and performed the MFM imaging measurements to directly probe the microscopic magnetic state. We found that the FM regions with typical sizes between 100 and 200 nm occupy about 50% of the entire film, while the rest of the film is non-FM down to the lowest achievable temperature of 4.5 K and up to the largest magnetic field of 13.4 T. The exchange bias measurements exclude the possibility of the non-FM phase to be antiferromagnetic. Preformed FM droplets were observed when temperature elevated 20 K above $T_C$, which indicates the existence of a Griffiths phase. Our study demonstrated the exotic sub-micron scale magnetic phase separation in high quality $LaCoO_3$ ultrathin films, which may have major implications in future oxide based electronics and spintronics.

## ACKNOWLEDGEMENTS

The authors thank Dr. Alexander J. Grutter for helpful discussions. This work was supported by the National Natural Science Foundation of China (Grants No. 51627901 and No. 11574287 and 11404300), Science Challenge Project (TZ2016003-1), National Key Research and Development Program of China (Grants No. 2016YFA0401004 and No. 2017YFA0402903), Hefei Science Center-Chinese Academy of Sciences (Grants No. 2016HSC-IU004), Anhui Initiative in Quantum Information Technologies (AHY100000). X.Z. acknowledges the support of the Youth Innovation Promotion Association CAS (Grant No. 2016389). The synchrony X ray measurements were performed at Beamline BL6.3.1.2 of the Advanced Light Source (Contract no. DE-AC02-05CH11231), 1W1A and 1W1B at Beijing Synchrotron Radiation Facility (BSRF), BL12B-a at the National Synchrotron

Radiation Laboratory (NSRL) of USTC and BL14W1 of the Shanghai Synchrotron Radiation Facility (SSRF).

## APPENDIX: METHODS

*Sample preparation and surface morphology characterizations.* The $LaCoO_3$ films with different thicknesses were deposited on $TiO_2$-terminated $SrTiO_3$ (0 0 1) substrates using pulsed laser deposition (PLD) monitored by reflection high-energy electron diffraction (RHEED). Before deposition, the $SrTiO_3$ substrates were chemically treated in buffered Hydrofluoric acid and then annealed at 930 °C in an oxygen atmosphere, to create atomically smooth surfaces with one-unit-cell-high terraces. The $LaCoO_3$ films were deposited using a laser fluency of 2 $J/cm^2$ and the frequency was set at 1 Hz. The film growth temperature and the oxygen partial pressure were 750 °C and 25 Pa, respectively. The surface morphology was characterized using a Bruker Multimode AFM system or Zeiss SEM system.

*Structural and Co valence characterizations.* Detailed information of XRD, XAS, XAFS and STEM measurements could be found in a previous paper [31]. The Synchrotron XRD measurements were carried out on BL14B1 at Shanghai Synchrotron Radiation Facility. The RSM were measured using a Rigaku Smartlab high resolution XRD. The Co *L*-edge XAS measurements were performed on BL6.3.1.2 at the Advanced Light Source (ALS) of Lawrence Berkeley National Laboratory (LBNL) and BL12B-a at the National Synchrotron Radiation Laboratory (NSRL) of USTC. The Co *K*-edge XAFS measurements were carried out on BL14W1 of SSRF in China. The valence calculation follows the same method as in reference [31]. The spherical-aberration corrected HAADF-STEM images and EELS results were acquired on a JEOL ARM200F microscope operating at 200 kV. The specimens were prepared using focused ion beam (FIB) along the pseudocubic (100) direction. In our STEM-HAADF measurements, we carefully avoided elongated measurements. Our HAADF images with typical size of 1024×1024 pixels were taken with a speed of 18 μs/pixel and the total exposure time to the electron beam (200 keV) is less than 20 sec.

*Bulk magnetization measurements.* The in-plane and out-of-plane magnetic measurements were performed in a Quantum Design SQUID vibrating sample magnetometer system. Quartz specimen holders with the minimum background signal were used to achieve a super high sensitivity, up to $10^{-9}$ emu. The M-H loops were obtained with an applied magnetic field from -3 to 3 T. The exchange bias measurements were done by measuring M-H loops after $\pm 3$ T field cooling to low temperatures. The M-T curves were obtained with an applied field of 500 Oe when the temperature was increasing after field cooling to the lowest temperature.

*MFM measurements and statistical analyses.* A home-built MFM system that can be inserted into a 20 T superconducting magnet was used to measure the topography and magnetic images. The details of the instrument can be found in Ref. [12,49]. In brief, the interaction between the magnetically coated tip and local magnetic field gradient of the sample would change the oscillation frequency of cantilever, which is recorded and used to image. The magnetic force between the tip and FM phase is attractive, leading to a negative frequency shift (blue color contrast) in MFM images. The weak or non-magnetic regions have negligible force on the tip, leading to a zero frequency shift (yellow color contrast) in MFM images. The external magnetic field is always parallel to oscillating direction of the MFM cantilever. The topography of the sample was imaged with the frequency-modulated tapping mode, and then the tip is lifted by about 100 nm to obtain the magnetic signal. All the analyses of MFM images, including FM area-percentage statistics, RMS frequency calculation and FM size distribution were performed with the aid of Gwyddion software. And the watershed algorithm implemented in Gwyddion software was used to obtain the FM area-percentage and size-distribution statistic results [50,51].

[37] H. Hsu, P. Blaha, and R. M. Wentzcovitch, Phys. Rev. B **85**, 140404 (2012).

[38] H. Seo, A. Posadas, and A. A. Demkov, Phys. Rev. B **86**, 014430 (2012).

[39] M. A. Torija, M. Sharma, J.Gazquez, M. Varela, C. He, J. Schmitt, J. A. Borchers, M. Laver, S. El-Khatib, and C. Leighton, Adv. Mater. **23**, 2711(2011)

[40] Y.-M. Kim, J. He, M. D. Biegalski, H. Ambaye, V. Lauter, H. M. Christen, S. T. Pantelides, S. J. Pennycook, S. V. Kalinin, and A. Y. Borisevich, Nat. Mater. **11**, 888 (2012).

[41] S. Park, P. Ryan, E. Karapetrova, J. W. Kim, J. X. Ma, J. Shi, J. W. Freeland, and W. Wu, Appl. Phys. Lett. **95**, 072508 (2009).

[42] C. Chen, J. Avila, E. Frantzeskakis, A. Levy, and M. C. Asensio, Nat. Commun. **6**, 8585 (2015).

[43] Y. Kim, J. He, M. Biegalski, H. Ambaye, V. Lauter, H. Christen, S. Pantelides, S. Pennycook, S. Kalinin, and A. Borisevich, Nat. Mater. **11**, 889 (2012).

[44] L. Zhang, C. Israel, A. Biswas, R. L. Greene, and A. de Lozanne, Science **298**, 805 (2002).

[45] R. B. Griffiths, Phys. Rev. Lett. **23**, 17 (1969).

[46] M. B. Salamon, P. Lin, and S. H. Chun, Phys. Rev. Lett. **88**, 4, 197203 (2002).

[47] S. J. May, P. J. Ryan, J. L. Robertson, J.-W. Kim, T. S. Santos, E. Karapetrova, J. L. Zarestky, X. Zhai, S. G. E. te Velthuis, J. N. Eckstein, S. D. Bader, and A. Bhattacharya, Nat. Mater. **8**, 892 (2009).
21

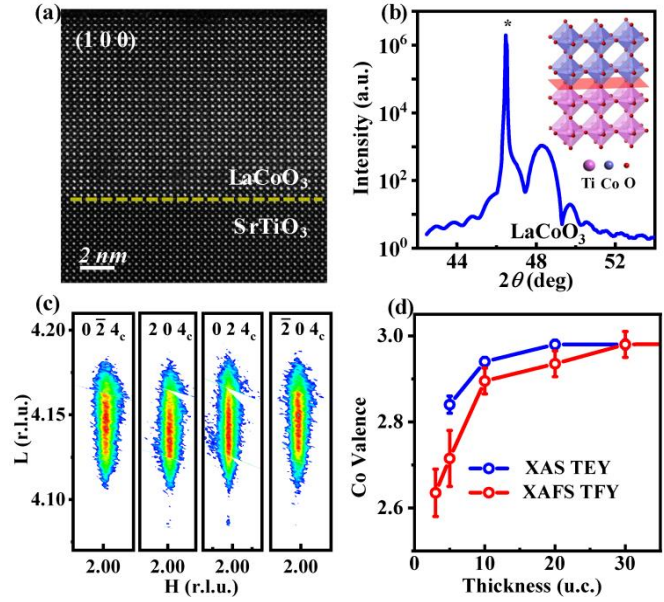

FIG. 1. (a) The HAADF-STEM image along the [100] direction. The yellow dashed line indicates the interface; (b) The XRD scan near the (0 0 2) peak; The star indicates the substrate. The schematic cartoon of the heterostructure with Sr and La sites omitted for clarity; (c) The XRD RSM of the LaCoO$_3$ film; (d) Co valences of different films measured by Co $L$-edge XAS TEY and $K$-edge XAFS TFY.

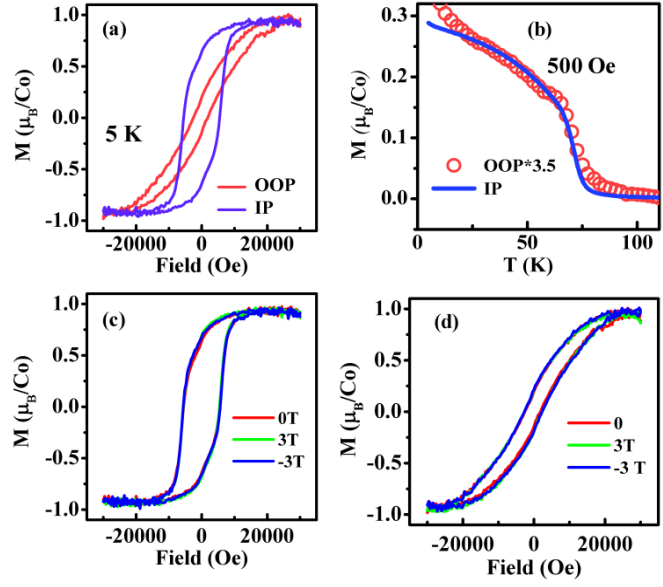

FIG. 2. (a) The in-plane (IP) and out-of-plane (OOP) M-H loops measured at 5 K and (b) M-T curves measured with an applied field of 0.05 T. The magnetic exchange bias measurements performed at 5 K along both (c) in plane and (d) out of plane directions.

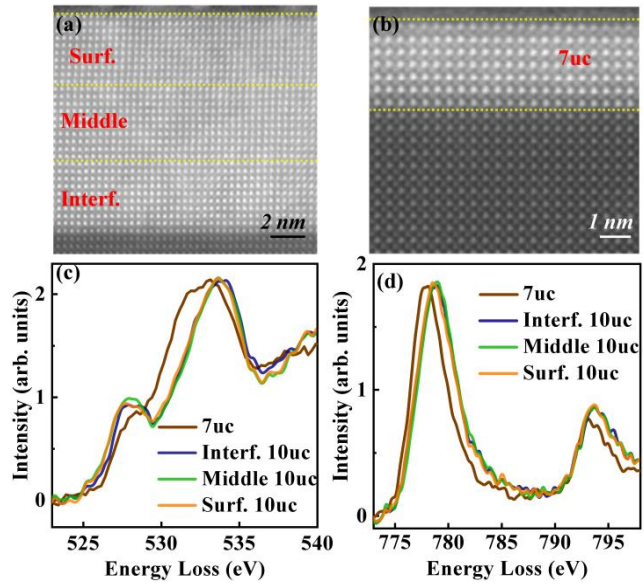

FIG. 3. The STEM images of the 30 unit cell LaCoO$_3$ film (a) and a 7 unit cell LaCoO$_3$ film (b). The average oxygen $K$-edge (c) and cobalt $L$-edge (d) in interface, middle and surface regions with 10 unit cells in each part, together with the average spectra of the 7 unit cell film.

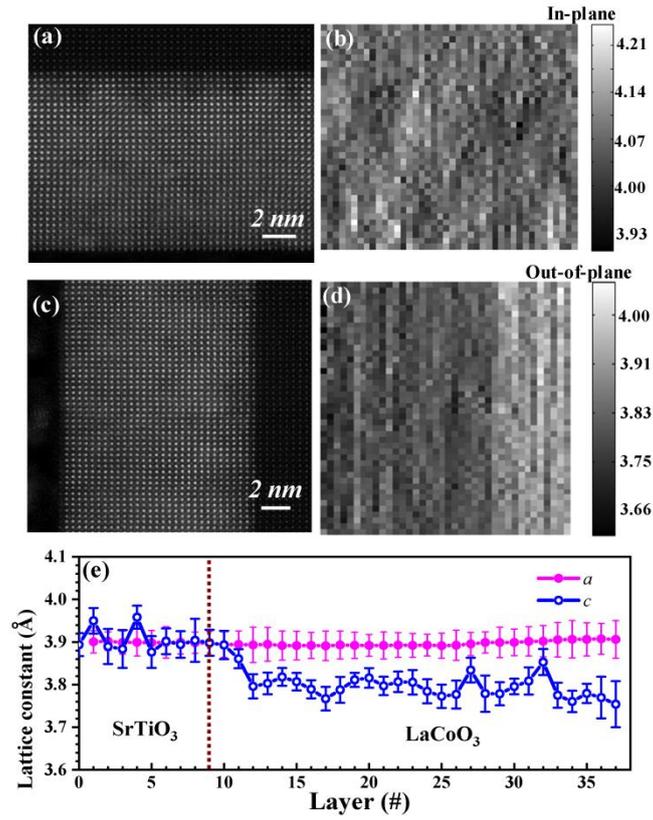

FIG. 4. (a)-(d), the STEM-HAADF image (left) and the quantized lattice constant (right) of the LaCoO$_3$ film on the SrTiO$_3$ substrate. (a),(c) are images scanned along the in-plane and out-of-plane directions respectively. (e), the quantized in-plane and out-of-plane lattice constants of each layer extracted from the bright-spot (La or Sr) distance in (b) and (d).

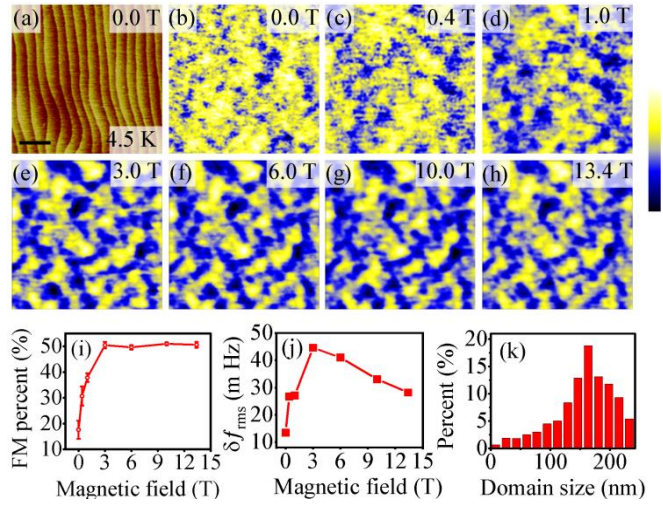

FIG. 5. (a) The topography image and (b)-(h) MFM images taken with increasing field. All images were taken at the same spot after zero field cooling; (i) FM area percentage and (j) the RMS of MFM frequency signal during the field increasing process; (k) The histogram of the FM domain size distribution of the MFM image shown in (e). For all images, the scanning area is 2×2 μm$^2$. The scale bar in (a) is 400 nm. And the color bars for (b)-(h) are 100, 175, 181, 284, 271, 216, and 199 mHz, respectively.

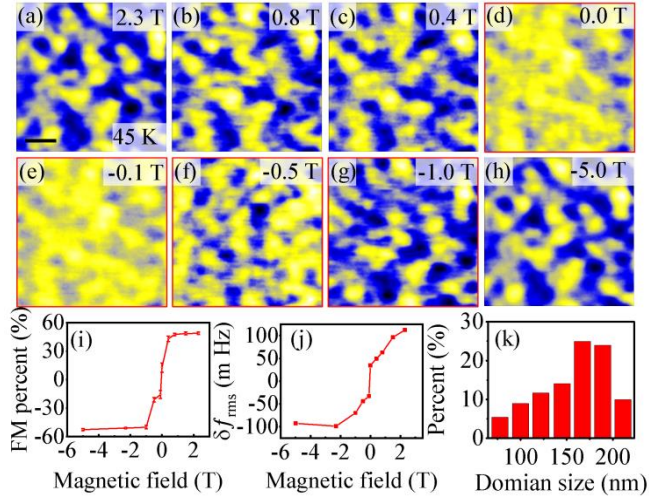

FIG. 6. MFM images taken at 45 K and the statistical analyses. (a)-(h) MFM images taken at the same location at 45 K after the 2.3 T field cooling. The FM spin flipping process occurs in the red-squared images. The scan area is 2×2 μm² for all images. The scale bar in (a) is 400 nm. And the color bars for (a)-(h) are 739, 400, 336, 215, 262, 285, 454, and 641 mHz, respectively. (i) The FM percentage and (j) RMS value of the MFM images as a function of the applied field shown in (a)-(h). (k) The histogram of the FM domain size distribution of the image shown in (a).

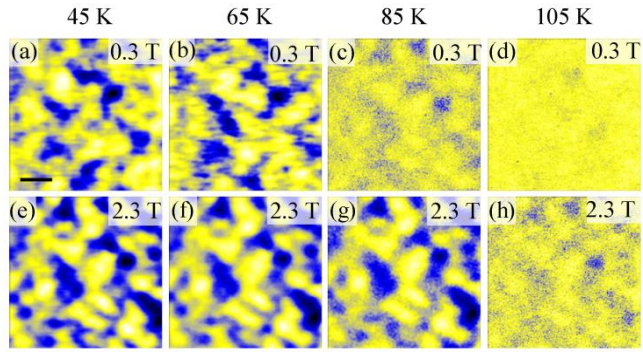

FIG. 7. (a)-(d) MFM images measured with applied fields of 0.3 T and (e)-(h) 2.3 T, respectively. The sample was first cooled down with the same fields prior to the MFM measurements. The scan size is 2×2 μm² for all images and all scans were taken at the same location of the film. The scale bar in (a) is 400 nm and color bars for (a)-(h) are 377, 217, 139, 86, 636, 410, 113, and 88 mHz, respectively.

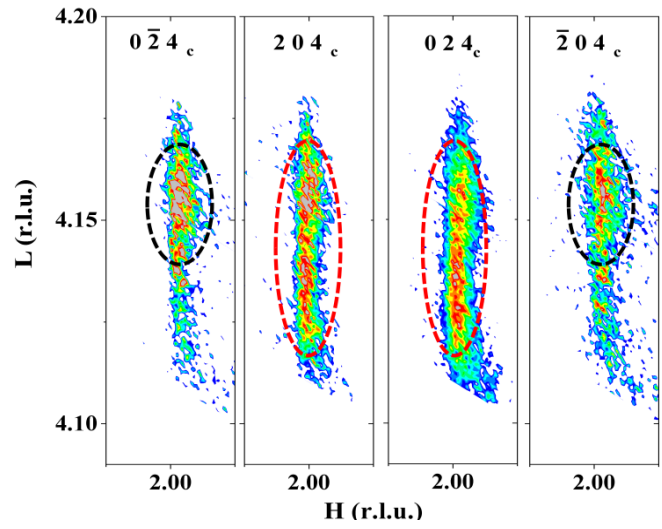

FIG. 8. The (2 0 4) XRD RSMs of a 40 unit cell film. The diffraction spots are diffusive and there are differences in the peak positions along the *L* direction as indicated by the black and red circles.

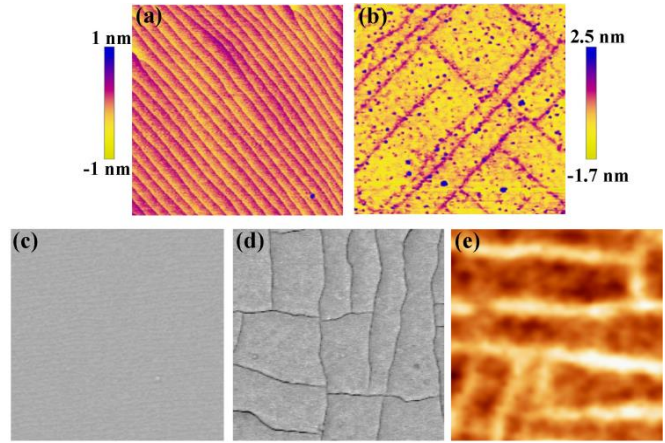

FIG. 9. Surface morphology differences between the 30, 90 and 180 unit cell films. (a) Atomic force microscopy images of a 30 unit cell film and (b) a 90 unit cell film. (c) The scanning electron microscopy image of the 30 unit cell film and (d) a 180 unit cell film. (e) The MFM results of the 180 unit cell film were measured at 7 K and 1.2 T. The cross-hatch lines observed in (b), (d) and (e) are similar to those in the reference [41], but are absent in the 30 unit cell film. The image area is 4×4 μm$^2$ in (a) and 5×5 μm$^2$ in (b)-(d).